\documentclass[useAMS,usenatbib]{mn2e}
\usepackage{natbibmnfix}
\usepackage{graphicx}
\usepackage{grffile}
\usepackage{subfigure}
\usepackage{mathrsfs,amssymb}
\usepackage{amsmath}
\usepackage{natbib}
\usepackage{color}
\usepackage{ulem}
\usepackage{url}
\usepackage{bm}

\newcommand{\araa}{\rm {ARA\&A}}             % Annual Review of Astron and Astrophys
\newcommand{\apj}{\rm {ApJ}}                 % Astrophysical Journal
\newcommand{\apjl}{\rm {ApJ}}                % Astrophysical Journal, Letters
\newcommand{\apjs}{\rm {ApJS}}               % Astrophysical Journal, Supplement
\newcommand{\aap}{\rm {A\&A}}                % Astronomy and Astrophysics
\newcommand{\jcap}{\rm {J. Cosmology Astropart. Phys.}}
\newcommand{\mnras}{\rm {MNRAS}}             % Monthly Notices of the RAS
\newcommand{\nat}{\rm {Nature}}              % Nature
\def\gtsima{$\; \buildrel > \over \sim \;$}
\def\ltsima{$\; \buildrel < \over \sim \;$}
\def\gsim{\lower.5ex\hbox{\gtsima}}
\def\lsim{\lower.5ex\hbox{\ltsima}}
\def\simgt{\lower.5ex\hbox{\gtsima}}
\def\simlt{\lower.5ex\hbox{\ltsima}}
\def\simpr{\lower.5ex\hbox{\prosima}}
 
 \def\CII{\hbox{C~$\scriptstyle\rm II$}}
 \def\CI{\hbox{C~$\scriptstyle\rm I$}} 
 \def\NII{\hbox{N~$\scriptstyle\rm II$}} 
 \def\OI{\hbox{O~$\scriptstyle\rm I$}}
    
 \newcommand*\oline[1]{%
  \vbox{%
    \hrule height 0.5pt%                  % Line above with certain width
    \kern0.25ex%                          % Distance between line and content
    \hbox{%
      \kern-0.1em%                        % Distance between content and left side of box, negative values for lines shorter than content
      \ifmmode#1\else\ensuremath{#1}\fi%  % The content, typeset in dependence of mode
      \kern-0.1em%                        % Distance between content and left side of box, negative values for lines shorter than content
    }% end of hbox
  }% end of vbox
}

\begin{document}
\title
[ {[\CII]} intensity mapping]
{Intensity mapping of [\CII] emission from early galaxies}

\author[Yue et al.]{B. Yue$^{1}$, A. Ferrara$^{1, 2}$,
 A. Pallottini$^{1}$, S. Gallerani$^{1}$, L. Vallini$^{1,3}$ \\
 $^1$Scuola Normale Superiore, Piazza dei Cavalieri 7, I-56126 Pisa, Italy\\
 $^2$Kavli IPMU (WPI), Todai Institutes for Advanced Study, the University of Tokyo, Japan\\
 $^3$Dipartimento di Fisica e Astronomia, Universit\'a di Bologna, viale Berti Pichat 6/2, 40127 Bologna, 
 Italy \\
 }

\maketitle
\begin{abstract}
The intensity mapping of the [\CII] 157.7~$\rm \mu$m fine-structure emission line represents an ideal experiment to probe star
formation activity in galaxies, especially in those that are too faint to be individually detected. Here, we investigate the feasibility of such an experiment for $z > 5$ galaxies. We construct the $L_{\rm CII} - M_{\rm h}$ relation from observations and simulations,
then generate mock [\CII] intensity maps by applying this relation to halo catalogs built from large scale N-body simulations. Maps of the extragalactic far-infrared (FIR) continuum, referred to as ``foreground", and CO rotational transition lines and [\CI] fine-structure lines referred to as ``contamination", are produced as well. We find that, at 316~GHz (corresponding to $z_{\rm CII} = 5$), the mean intensities of the extragalactic FIR continuum, [\CII] signal, all CO lines from $J=1$ to 13  and two [\CI] lines are $\sim 3\times10^5$~Jy~sr$^{-1}$, $\sim 1200$~Jy~sr$^{-1}$,  $\sim 800$~Jy~sr$^{-1}$ and $\sim 100$~Jy~sr$^{-1}$, respectively. We discuss a method that allows us to subtract the FIR continuum foreground by removing a spectrally smooth component from each line of sight, and to suppress the CO/[\CI] contamination by discarding pixels that are bright in contamination emission. The $z > 5$ [\CII] signal comes mainly from halos in the mass range $10^{11-12} ~M_\odot$; as this mass range is narrow, intensity mapping is an ideal experiment to investigate these early galaxies. In principle such signal is accessible to a ground-based telescope with a 6~m aperture, 150~K system temperature, a $128\times128$ pixels FIR camera in 5000 hr total integration time,
however it is difficult to perform such an experiment by using currently available telescopes.
\end{abstract}

\begin{keywords}
cosmology: diffuse radiation-dark ages; reionization, first stars -- radio lines: galaxies -- galaxies: high-redshift 
\end{keywords}

\section{introduction}

The emission line intensity mapping is a technique to access high-$z$ galaxies below the detection limit without losing redshift information, as proposed by e.g. \citet{2010JCAP...11..016V, 2011JCAP...08..010V, 2011ApJ...728L..46G, 2012ApJ...745...49G,2013ApJ...768..130G,2011ApJ...741...70L, 2013ApJ...763..132S,2014ApJ...786..111P}.
Optimistically, it only collects radiation from galaxies in a selected redshift range, as the spurious flux due to foregrounds, contaminating radiation and noise can be in principle removed or suppressed. 
Compared with galaxy surveys that aim at resolving faint spots in a limited field of view (FOV), the advantages of intensity mapping rely on the fact that, if the galaxy luminosity function has a sufficiently steep faint end, the observed radiation is actually dominated by unresolved sources \citep{2014ApJ...793..116U}. Even if this is not the case, intensity mapping can still be used to study unresolved galaxies once resolved sources are removed (masked). Interestingly, an intensity mapping experiment could be carried out with a modest aperture but large FOV telescope.

The [\CII] 157.7~$\mu$m fine-structure line arising from the $^2$P$_{3/2}$$\rightarrow$$^2$P$_{1/2}$ transition
is the brightest amongst all metal lines emitted by the interstellar medium (ISM) of star-forming galaxies.
It is associated to the star formation in galaxies \citep{2002A&A...385..454B,2011MNRAS.416.2712D,2014arXiv1402.4075D,2014arXiv1409.7123H} and plays a key role in the energy balance of galaxies, as it provides one of the most efficient cooling processes for the neutral ISM. With respect to the Ly$\alpha$ line, the [\CII] line has the advantage of being unaffected by dust attenuation and neutral hydrogen absorption.

In the local Universe, [\CII] line has been successfully detected even in galaxies with amazingly low star formation rates (SFR) of $\sim 0.001~M_\odot$yr$^{-1}$ \citep{2014arXiv1402.4075D}. These authors have also derived the relation between the [\CII] line luminosity, $L_{\rm CII}$, and the SFR of \textit{local} galaxy samples \citep{2014arXiv1402.4075D}. Surprisingly, given the rather complicated physics behind the [\CII] emission, $L_{\rm CII}$ scales rather tightly with SFR. However, at high redshift ($z \gsim4$), the [\CII] line has been detected so far only in quasar host galaxies \citep{2005A&A...440L..51M, 2012ApJ...751L..25V,2012A&A...543A.114G,2013ApJ...773...44W,2013ApJ...770...13W,2014arXiv1409.4418C} or ultra-luminous infrared galaxies (ULIRGs, with $L_{\rm IR} > 10^{12}~L_\odot$ where $L_{\rm IR}$ is the in-band luminosity at $8-1000~\mu$m) characterized by SFR $\sim 10 ^{2-3} ~M_\odot$yr$^{-1}$ \citep{2011ApJ...740...63C, 2011A&A...530L...8D, 2014A&A...565A..59D}.
For typical normal star-forming galaxies (SFR $\sim 10~M_\odot$yr$^{-1}$), [\CII] emission has not yet been detected \citep{2013ApJ...778..102O, 2014ApJ...792...34O,2014arXiv1407.5793S,2014ApJ...784...99G}. This might indicate that most of carbon in these galaxies is at higher ionization state and/or their ISM is 
characterized by a very low level of metal enrichment.
By applying the $L_{\rm CII} - $SFR relation derived from local galaxies samples to high redshift Ly$\alpha$ emitters it is possible to compute the expected [\CII] flux from these galaxies. The fact that their [\CII] line remains undetected even with ALMA provides useful constraints on their internal radiation field, molecular content, gas density, and metallicity \citep{V15,2013MNRAS.433.1567V,2014ApJ...784...99G}.

As probing tools, intensity mapping experiments are affected by the presence of foreground radiation, including that represented by the galaxy continuum redshifted into the observed band.
Unfortunately, it is almost often the case that the foreground intensity largely exceeds that of the signal. The typical [\CII] line luminosity is $0.1\% - 1\%$ of the $L_{\rm IR}$ \citep{2009A&A...500L...1M}.
This implies that even if only one percent of the IR luminosity is redshifted into the observed band, the continuum emission overcomes the [\CII] line.

In addition to the far-infrared (FIR) continuum foreground, there are other emission lines emitted from a range of redshifts that fall at the same frequency of the [\CII] signal; they act as \textit{contaminants}.
 For example, the [\OI] line with wavelength 145~$\mu$m, the two [\NII] lines ($\lambda = 122, 205~\mu$m) and two [\CI] lines ($\lambda= 610, 371~\mu$m), and a handful of CO rotational transition lines in the range 200-2610 $\mu$m. Among these, the CO rotational transition lines are the most relevant here. For example, since the CO(4-3) line has a wavelength $651~\rm \mu m$, if emitted from $z = 0.45$ galaxies, it contaminates the [\CII] emission from $z = 5$ galaxies. The emission efficiency\footnote{As a caveat, we note that there is no clear consensus in the literature on this value, see \citet{2014MNRAS.443.3506B}.} of the CO(4-3) line from star-forming galaxies is $\sim 2\%$ of the [\CII] line \citep{2010JCAP...11..016V}. However, the luminosity distance from $z=0$ to 0.45 is only $\sim 5\%$ of that to $z=5$. As the flux is inversely proportional to the square of the luminosity distance, whereas the proper distance interval that corresponds to the same bandwidth is $\propto (1+z)^{-3/2}$, the CO flux can be more than ten times higher than the [\CII] one, even ignoring the cosmological evolution of the star formation rate density. Thus, CO contamination, as well as the continuum foreground, cannot be ignored and must be considered thoroughly. 
 
Although the [\CII] signal itself can be computed analytically \citep{2012ApJ...745...49G,2014ApJ...793..116U}, a reliable investigation of the influence of foreground/contamination is only possible based on mock maps that carefully mimic observations as close as possible. This is the prime motivation of this paper. Using halo catalogs recovered from large scale N-body simulations, we produce mock maps including (a) [\CII] signal, (b) FIR continuum foreground, (c) CO and [\CI] contamination lines, and (d) instrumental noise. We then test our foreground/contamination removal scheme on these maps to demonstrate the successful recovery of the original [\CII] signal.

The layout of this paper is as follows. In Sec. \ref{method}, we describe the model used to compute the [\CII] from high-$z$ galaxies and the necessary steps to generate mock maps. We show our forecasts for the [\CII] signal, extragalactic FIR continuum and the contamination and perform foreground/contamination removal experiments on mocks to recover the original [\CII] signal. The results are presented in Sec. \ref{result}. The conclusions and discussion are found in Sec. \ref{conclusion}.

\section{methods}\label{method}

\subsection{[\CII] emission from early galaxies}\label{obs}

\citet{2013MNRAS.433.1567V} and \citet{V15} (hereafter V15) have combined high-$z$ galaxy numerical simulations with sub-grid models of the ISM to compute the expected [\CII] luminosity ($L_{\rm CII}$) arising from
diffuse neutral gas and photodissociation regions (PDRs).
The resulting trend of $L_{\rm CII}$ with SFR and metallicity ($Z$) is 
consistent with observations of local metal-poor dwarf galaxies \citep{2014arXiv1402.4075D}. 
For the range [0.1,100]~$M_\odot$yr$^{-1}$ and [0.05,1.0]~$Z_\odot$
which should encompass most of the sources contributing to the total [\CII] emission at high redshift, V15 results are well reproduced by the following fitting formula
\begin{align}
{\rm log}(L_{\rm CII})& = 7.0 + 1.2\times{\rm log(SFR)} + 0.021\times{\rm log(Z)} \nonumber \\
&+ 0.012\times{\rm log(SFR)log(Z)} - 0.74\times{\rm log^2(Z)},
\label{LCII}
\end{align}
where $L_{\rm CII}$, SFR and $Z$ are in units of $L_\odot$, $M_\odot$yr$^{-1}$ and $Z_\odot$ respectively. 

The next step is to compute the $L_{\rm CII}-M_{\rm h}$ relation, where $M_{\rm h}$ is the halo mass.
To this aim we need to know the SFR$-M_{\rm h}$ and $Z-M_{\rm h}$ relations.
We obtain them from the procedure described in below paragraphs\footnote{Note that, compared to galaxies, the intergalactic medium emits a negligible [\CII] signal (e.g. \citealt{2012ApJ...745...49G}) at $z\gsim2$, and therefore is not considered in this work.}.

Since the UV luminosity ($L_{\rm UV}$) of a galaxy scales with its SFR
(e.g. \citealt{1998ARA&A..36..189K}), we adopt the observed UV luminosity functions (LFs)
to derive the SFR$-M_{\rm h}$ relation.
The {\it measured} UV LF is well described by a Schechter parameterization \citep{1976ApJ...203..297S}: 
\begin{equation}
\frac{dn}{dM_{\rm UV}} = 0.4\,{\rm ln}(10)\,\phi_\star \, x^{1+\alpha} e^{-x},
\label{dndMUV}
\end{equation}
where $x= 10^{0.4(M^\star_{\rm UV}-M_{\rm UV})}$, with $M_{\rm UV}$ the dust-attenuated absolute magnitude. For the rest-frame UV luminosity at 1600 \AA, the redshift-dependent parameters $(M^\star_{\rm UV}, \phi_\star, \alpha)$ that fit observations between $z\sim4-8$ are \citep{2014arXiv1403.4295B}
\begin{align}
M^\star_{\rm UV} &= -20.96+0.01(z-6) \nonumber \\
\phi_\star&=0.46\times10^{-0.27(z-6)}10^{-3} \\
\alpha&=-1.87-0.10(z-6)\nonumber\,.
\end{align}
The intrinsic absolute magnitude is $M^\prime_{\rm UV}=M_{\rm UV}-A_{1600}$, where $A_{1600} = 4.43+1.99\beta$ ($\ge 0$) is the dust attenuation at 1600 \AA ~\citep{1999ApJ...521...64M} and $\beta$ is the measured spectral slope ($f_\lambda\propto \lambda^\beta$).
Normally $\beta$ depends on $M_{\rm UV}$ and is fitted by \citep{2014ApJ...793..115B}
\begin{equation}
\beta = \beta_{-19.5}+\frac{d\beta}{dM_{\rm UV}}(M_{\rm UV}+19.5).
\end{equation}
From Fig. 2 of \citet{2014ApJ...793..115B} we find the following redshift-dependent fit, valid for $4 \simlt z\simlt 7$ 
\begin{align}
&\beta_{-19.5}=-1.97-0.06(z-6) \nonumber \\
&\frac{d\beta}{dM_{\rm UV}}=-0.18-0.03(z-6).
\end{align}
The intrinsic UV LF is then connected to the measured UV LF via 
\begin{equation}
\frac{dn^\prime}{dM^\prime_{\rm UV}}(M^\prime_{\rm UV},z)=\frac{dn}{dM_{\rm UV}}(M_{\rm UV},z).
\end{equation}

Assuming that the intrinsic $L^\prime_{\rm UV}$ monotonically increases with $M_{\rm h}$ and that all halos host some star formation activity, we obtain the $L^\prime_{\rm UV}-M_{\rm h}$ relation from 
\begin{equation}
\int_{M^\prime_{\rm UV}} \frac{dn^\prime}{d{M^\prime_{\rm UV}}} d{M^\prime_{\rm UV}} = \int_{M_{\rm h}} \frac{dn}{dM_{\rm h}}dM_{\rm h},
\end{equation}
where $dn/dM_{\rm h}$ is the halo mass function \citep{1999MNRAS.308..119S,2001MNRAS.323....1S}. This ``abundance matching" technique will be used also in Sec. \ref{FIRcontinuum} to derive the relation 
between the IR luminosity and halo mass.
We then derive the SFR from $L^\prime_{\rm UV}$. In principle $L^\prime_{\rm UV}$ depends not only on 
the SFR, but also on metallicity and stellar age. However, we note that the UV luminosity is insensitive to the metallicity and the stellar age, unless stars are very young ($\lsim 10$~Myr). So we can safely assume that $L^\prime_{\rm UV}$ scales with SFR as 
\begin{equation}
L^\prime_{\rm UV} =l_{\rm UV} \times {\rm SFR}.
\end{equation} 
We compute $l_{\rm UV}$ from {\tt Starburst99}\footnote{http://www.stsci.edu/science/starburst99/docs/default.htm} \citep{1999ApJS..123....3L,2005ApJ...621..695V,2010ApJS..189..309L}
by assuming a metallicity $0.1~Z_\odot$, stellar age 10\% of Hubble time, and a Salpeter IMF between $0.1- 100~M_\odot$. We choose the ``continuous star formation" mode. At 1600 \AA, $l_{\rm UV} =(8.9, 8.6, 8.3)\times10^{27}$~erg s$^{-1}$Hz$^{-1}$($M_\odot $/yr$)^{-1}$ at $z= (5, 6, 7)$; which is similar to \citet{1998ARA&A..36..189K}.

\begin{figure}
\centering{
\includegraphics[scale=0.4]{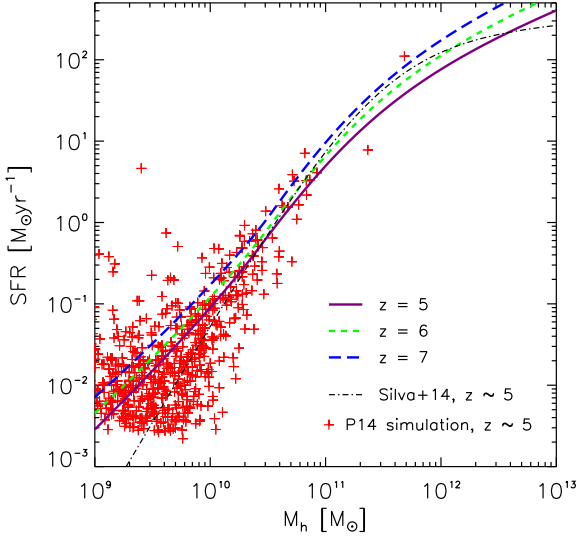}
 \caption{The SFR derived from UV LFs as a function of halo mass at $z = 5,\, 6\,{\rm and}\, 7$ respectively. The SFR$-M_{\rm h}$ relation in \citet{2014arXiv1410.4808S} at $z\sim5$ and the SFR of each halo containing Pop II stars in P14 simulation at $z\sim5$ are also shown.}
\label{SFR_M}
}
\end{figure}

We plot the SFR derived with this procedure at various redshifts as a function of $M_{\rm h}$ in Fig. \ref{SFR_M}. As a comparison we also plot the SFR$-M_{\rm h}$ relation at $z\sim5$ found by \citet{2014arXiv1410.4808S} who fit the relation using semi-analytical models of galaxy formation.
As can be seen by inspecting this figure, the specific SFR, i.e. the SFR per unit mass, starts to drop at a turnover mass $\sim10^{11}~M_\odot$. This is consistent with semi-analytical model predictions.

\begin{figure}
\centering{
\includegraphics[scale=0.4]{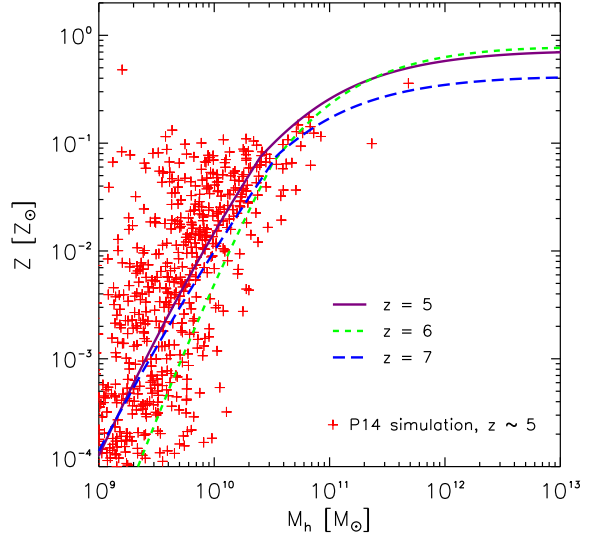}
 \caption{Metallicity, derived from mass-to-light ratio and the FMR, as a function of halo mass at $z = 5,\, 6\,{\rm and}\, 7$. The metallicity of each halo containing Pop II stars in the P14 simulation at $z\sim5$ are also shown.
}
\label{Z_M}
}
\end{figure}

The final ingredient of Eq. (\ref{LCII}) is $Z$. As the metallicity of high-$z$ galaxies is very 
poorly constrained at present, we derive it by combining the $L^\prime_{\rm UV} - M_{\rm h}$ relation
with the ``fundamental metallicity relation" (FMR) that relates $Z$ to the stellar Mass ($M_\star$) 
and SFR. The FMR inferred from low-$z$ galaxy observations \citep{2010MNRAS.408.2115M} 
is given by the following equation:
\begin{align}
\log(Z) &= 0.21+0.37\log(M_{10})-0.14\log({\rm SFR})\nonumber\\
&-0.19\log^{2}(M_{10})-0.054\log^{2}({\rm SFR})\\
&+0.12\log(M_{10})\log({\rm SFR})\nonumber,
\end{align}
where $M_{10}=M_\star/10^{10}$, $M_\star$ and $Z$ are expressed in Solar units.
No redshift evolution is found at least up to $z= 2.5$ \citep{2010MNRAS.408.2115M}; therefore we apply it also to high-$z$ galaxies, with the caveat that deviations might appear for high-$z$ galaxies. The stellar mass, $M_\star$, is linked to the UV absolute magnitude via the mass-to-light ratio. From the latest measurements \citep{2014MNRAS.444.2960D}, 
\begin{equation}
{\rm log}(M_\star) ={\rm log}(M_\star^{0})+\frac{dM_\star}{dM_{\rm UV}}(M_{\rm UV}+19.5),
\end{equation}
where $\log(M_\star^{0}) = (9.00, 8.84, 8.63)$ and $dM_\star/dM_{\rm UV} = (-0.46, -0.54, -0.45)$ in the redshift ranges $4.5 \le z < 5.5$, $5.5 \le z < 6.5$, and $6.5 \le z < 7.5$, respectively, for the model without nebular line contribution. We use the observed UV magnitude $M_{\rm UV}$ (i.e., without dust correction), and we ignore the negligible difference between the UV luminosity at 1500 \AA~and 1600 \AA.
Fig. \ref{Z_M} shows the derived $Z-M_{\rm h}$ relation at $z=5, 6$ and 7.

\begin{figure}
\centering{
\includegraphics[scale=0.4]{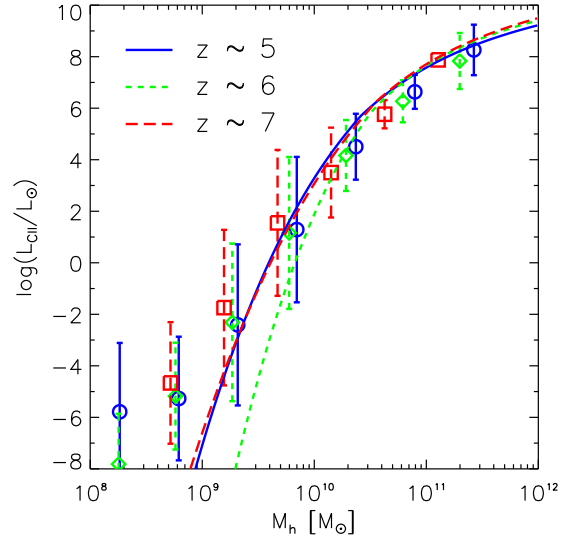}
 \caption{The mean $L_{\rm CII}$ as a function of halo mass, $M_{\rm h}$, at redshifts $\sim$5, 6 and 7 respectively. The lines are the relation derived using observations while the points are for the P14 simulation (the errorbars are the standard deviation of $\langle \log(L_{\rm CII}) \rangle$ for each mass bin).
 }
\label{logLCII_M}
}
\end{figure}

We finally compute the [\CII] luminosity of halos with mass $M_{\rm h}$ by substituting the SFR and $Z$ derived above into 
Eq. (\ref{LCII}).
Although different star formation histories may cause a scatter in the [\CII] luminosities for halos of a given mass, we neglect this effect because it only results in noise as long as the luminosity dispersion is independent of position on large scales. The $L_{\rm CII} - M_{\rm h}$ relation derived with the above procedure is shown in Fig. \ref{logLCII_M} by solid, short dashed and long dashed lines for $z=5, 6$ and 7 respectively.

When deriving the above $L_{\rm CII} - M_{\rm h}$ relation, the properties of faint galaxies are extrapolated from the observed bright galaxies. 
We check the validity of this relation by comparing the result obtained through this semi-empirical method with the numerical simulations of cosmic metal enrichment presented by \citet{2014MNRAS.440.2498P} (hereafter P14). P14 have used an hydrodynamical simulation to follow the star formation and the Pop III-Pop II transition ($Z > 10^{-4}~Z_\odot$). Hereafter, we only consider the Pop II star formation mode.

If a halo has formed stars, it contains a number of stellar particles whose birth date and metallicity is recorded. For a selected halo, the mean stellar age is 
\begin{equation}
t_{\rm age} = \frac{\sum_{i} \Delta t_i m_{\star,i}}{\sum_{i} m_{\star,i}},
\end{equation}
where $\Delta t_i$ is the birth date of the $i$-th stellar particle, $m_{\star,i}$ is its mass and the sum is extended over all Pop II stellar particles in that halo. The mean metallicity of the halo is 
\begin{equation}
Z = \frac{\sum_{i} Z_i m_{\star,i}}{\sum_{i} m_{\star,i}},
\end{equation}
where $Z_i$ is the metallicity of the $i$-th Pop II stellar particle. By dividing the total Pop II stellar mass by the mean age, we obtain the mean SFR of the halo,
\begin{equation}
{\rm SFR} = \frac{\sum_i m_{\star,i}}{t_{\rm age}}.
\end{equation}
The SFR vs. $M_{\rm h}$ and $Z$ vs. $M_{\rm h}$ at $z\sim5$ in P14 simulation are also plotted in Fig. \ref{SFR_M} and Fig. \ref{Z_M} respectively.

Using Eq. (\ref{LCII}) we calculate the [\CII] luminosity of halos from their SFR and $Z$ and group halos into several mass bins at each simulation output. Some halos only have Pop III stars, or are too small to host any star formation. Therefore in each mass bin only a fraction $f_{\rm CII}$ of halos exhibit [\CII] emission. This fraction tends to one as the halo mass increases. We denote the mean log of [\CII] luminosity for {\it halos that exhibit [\CII] emission} by $\langle \log (L_{\rm CII}) \rangle$, and use 
\begin{equation}
{\rm log}(L_{\rm CII}) = \langle{\rm log}(L_{\rm CII})\rangle+{\rm log}(f_{\rm CII})
\end{equation}
as the mean [\CII] luminosity of {\it all} halos with mass $M_{\rm h}$; this quantity is plotted in Fig. \ref{logLCII_M}.

\subsection{Far-infrared continuum foreground}\label{FIRcontinuum}

In this subsection, we model the extragalactic foreground due to the FIR continuum from galaxies at different redshifts. The Milky Way FIR and CMB radiation are assumed to be removed straightforwardly and hence are not considered in this work.

The FIR luminosity function of galaxies, including spiral galaxies, starburst galaxies, star-forming galaxies containing AGNs and sometimes AGNs, is studied in e.g. \citet{2009A&A...496...57M,2013MNRAS.432...23G,2013A&A...553A.132M}. In \citet{2013MNRAS.432...23G}, the LF can 
be written as
\begin{equation}
\Phi = \Phi_\star\left(\frac{L_{\rm IR}}{L^\star_{\rm IR}}\right)^{1-\alpha}
{\rm exp}\left[-\frac{1}{2\sigma^2}{\rm log}^2\left(1+\frac{L_{\rm IR}}{L^\star_{\rm IR}}\right)\right],
\label{PhiIR}
\end{equation}
where $L_{\rm IR }$ is the infrared luminosity between $8 - 1000~{\rm \mu}$m.
We use redshift evolution formulae of parameters $\alpha, \sigma, \Phi_\star, L_{\rm IR}^\star$
\citep{2013MNRAS.432...23G}: $\Phi_\star = 5.7\times10^{-3}(1+z)^{-0.57}$ for $z \le 1.1$, and $\Phi_\star = 6.81\times10^{-2}(1+z)^{-3.92}$ for $z > 1.1$; $L_{\rm IR}^\star = 7.68\times10^9 (1+z)^{3.55}$ for $z \le 1.85$ and $L_{\rm IR}^\star =5.80\times10^{10}(1+z)^{1.62}$ for $z > 1.85$; $\alpha = 1.15$, $\sigma = 0.52$ for $z \le 0.3$, and $\alpha = 1.2$, $\sigma = 0.5$ otherwise. The above LFs are constructed from galaxy samples at $z\simlt 4.2$, therefore including the large majority of the sources contributing to the FIR continuum (and CO, which is associated to the FIR continuum, see next subsection).

We use again the abundance matching technique \citep{2012A&A...537L...5B} to construct the $L_{\rm IR} - M_{\rm h}$ relation. We suppose that the contribution of subhalos to the IR luminosity function is small and we ignore them. By equating the number density of galaxies with IR luminosity above $L_{\rm IR}$ and the number density of halos above $M_{\rm h}$, 
\begin{equation}
\int_{L_{\rm IR}} \Phi(L_{\rm IR},z)dL_{\rm IR} = \int_{M_{\rm h}} \frac{dn}{dM_{\rm h}}dM_{\rm h},
\label{LIR}
\end{equation}
the $L_{\rm IR} - M_{\rm h}$ relation is derived. 

\begin{figure}
\centering{
\includegraphics[scale=0.4]{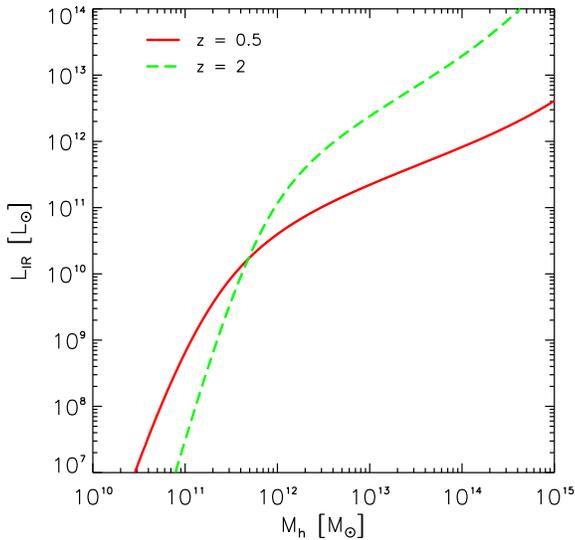}
\caption{FIR continuum luminosity derived from Eq. (\ref{LIR}) as a function halo mass at redshift $z=0.5$ and 2.
 }
\label{L_IR}
}
\end{figure}

We plot the the IR luminosity - halo mass relation at redshift 0.5 and 2 in Fig. \ref{L_IR}. For the same reasons given in Sec. \ref{obs}, we do not consider the IR luminosity dispersion among halos with the same mass $M_{\rm h}$. 
 
\subsection{CO and [\CI] emission lines contamination}

The CO rotational transition lines from low-redshift galaxies are by far the most important contaminants for the $z>5$ \CII signal. Although some studies aimed at measuring CO luminosity functions exist (e.g. \citealt{2003ApJ...582..659K} and references therein), they lack data either for higher rotational transition numbers, $J$, or at high redshift. On the other hand, the CO line luminosity is found to be closely related to the IR luminosity, since both lines are good star formation activity tracers \citep{2009MNRAS.399..264B,2014MNRAS.444.1301P}.

The CO line luminosity is derived by using the ${\rm log}(L_{\rm IR}) =\alpha{\rm log}(\oline{L}_{\rm CO})+\beta$ relations presented in Tab. 3 of \citet{2014ApJ...794..142G} for lines with $J$-ladders from 1 to 13. These relations are fitted from samples of local ($z < 0.1$) (U)LIRGs and high-$z$ ($z > 1$) dusty star-forming galaxies (DSFGs). Since CO lines are considered as contaminants to be removed when recovering the [\CII] signal, the scatter around the fitting relation should be modeled. To this aim we assume a Gaussian distribution of the form 
\begin{equation}
p(L_{\rm CO}|~\oline{L}_{\rm CO}) =\frac{1}{\sqrt{2\pi} \sigma} {\rm exp} 
\left[-\frac{x^2}{2\sigma^2}\right],
\label{pCO}
\end{equation}
where $x= {\rm log}( L_{\rm CO}) - {\rm log}( \oline{L}_{\rm CO} )$ and $\sigma = s/\alpha$ is the variance; 
$s$ is the scatter of the data around the ${\rm log}(L_{\rm IR}) - {\rm log} (\oline{L}_{\rm CO}) $ fitting in \citet{2014ApJ...794..142G}, including the intrinsic dispersion and statistical errors. 

We further consider the contamination from two [\CI] fine-structure lines: (i) [\CI(1-0)], corresponding to the $^3$P$_{1}$$\rightarrow$$^3$P$_{0}$ transition, at 492~GHz, and (ii) and [\CI(2-1)], corresponding to the $^3$P$_{2}$$\rightarrow$$^3$P$_{1}$ transition, with frequency 809~GHz. Several authors have reported [\CI] observational data (e.g. \citealt{2000ApJ...537..644G, 2002A&A...383...82I,2011ApJ...730...18W,2013MNRAS.435.1493A}), finding relations between the [\CI] and CO or IR luminosities. Motivated by observations, \citet{2014MNRAS.444.1301P}  have calculated the expected [\CI]/$L_{\rm IR}$ ratios for $0 < z < 2$ galaxies, by combining a semi-analytical galaxy formation model with radiative-transfer and line-tracing calculations. We adopt the outcome of these theoretical calculations and we add 0.25 dex scatters to the mean ratio, namely the maximum of the deviations reported by \citet{2014MNRAS.444.1301P}.

\subsection{Instrumental noise}\label{sec_instrumental} 
We have to account for instrumental noise, in order to have predictions that can be fairly compared with observation.
The noise level of a radio telescope is given by the standard expression 
\begin{equation}\label{eq_noise}
\sigma_{\rm N} = \frac{2k_BT_{\rm sys}}{A\sqrt{\Delta\nu_0 t}},
\end{equation}
where $k_B$ is the Boltzmann constant, $T_{\rm sys}$ is the system temperature, $A$ is the area of the antenna, $t$ is the integration time per FOV. If the camera has $N_{\rm pix}$ pixels and the observation is performed at wavelength $\lambda_0$, normally the instrument is designed by such a way that ${\Omega_{\rm FOV}} \sim \lambda_0^2/AN_{\rm pix}$ holds. To cover a sky region of solid angle $\Omega_{\rm map}$, the total integration time is $t_{\rm obs} = t\times \Omega_{\rm map}/\Omega_{\rm FOV}$. We model the instrumental noise as a zero mean Gaussian random variable without spatial and frequency correlation, and we add such fluctuations to the mock maps.
 
\subsection{Mock maps}

The light cone for which we produce the intensity maps is built from the halo catalogs of the 
{\tt BolshoiP} simulation\footnote{\url{http://www.cosmosim.org/cms/simulations/bolshoip-project/bolshoip/}}
({\tt Bolshoi} simulation with {\tt Planck} cosmology, see {\tt Bolshoi} simulation paper \citealt{2011ApJ...740..102K}).
In the simulation, the smallest halos resolved are $\approx5\times10^9~M_\odot$, well below the mass of halos that are expected to host the bulk of [\CII] emission (see Sec.s \ref{COmasking}).

The box of this simulation is $L=250~h^{-1}$cMpc on a side, corresponding to 2.4 degree when located at $z=7$.
When making light cones from a simulation with a box length smaller than the cone depth, a standard protocol is to replicate the same halo catalog along the radial direction at the same time applying a ``randomization'' procedure made of random translations, rotations and reflections in order to avoid spurious periodicity effects \citep{2005MNRAS.360..159B}.

We divide a $2.4\times2.4$~deg$^2$ sky region into $200\times200$ pixels whose angular size is $43''$. This corresponds to the beam size of a 6~m radio telescope for $\lambda_0=\lambda_{\rm CII }(1+7)$. The frequency range [238, 317]~GHz ($z_{\rm CII} = $[7, 5]) is equally divided into 60 bins with bandwidth of each bin $\Delta\nu_0 = 1.3$~GHz.

Given a pixel in the [\CII] map, the measured intensity in the frequency bin centered at $\nu_0$ is
\begin{equation}
I_{\rm CII}(\nu_0) = \frac{1}{(\Delta \theta)^2}\sum_j \frac{1}{\Delta \nu_0}\frac{L^j_{\rm CII}}{4\pi r_j^2(1+z_j)^2},
\label{ICII}
\end{equation}
where $\Delta \theta$ is the angular size of the pixel and $r_j$ is the comoving distance up to $z_j$; the sum is performed on all halos seen by this beam and with redshift 
\begin{equation}
\frac{\nu_{\rm CII}}{\nu_0+\Delta \nu_0/2}-1 \le z_j \le\frac{\nu_{\rm CII}}{\nu_0-\Delta \nu_0/2}-1.
\label{zj}
\end{equation}

For the CO and [\CI] emission lines the procedure is the same as for [\CII]. For the FIR continuum, Eq. (\ref{ICII}) becomes
\begin{equation}
I_{\rm FIR}(\nu_0)=\frac{1}{(\Delta \theta)^2}\sum_j \frac{L^j_{\rm IR}{\rm SED}_{\rm IR}(\nu)(1+z_j)}{4\pi r_j^2(1+z_j)^2},
\end{equation}
where $\nu = \nu_0(1+z_j)$, and ${\rm SED}_{\rm IR}(\nu)$ is the normalized spectrum template of the galaxy, 
\begin{equation}
\int_{c/1000\mu m}^{c/8\mu m} {\rm SED}_{\rm IR}(\nu) d\nu = 1.
\end{equation}
 For simplicity, we choose the Spi4 spiral galaxy SED template from the {\tt SWIRE} template library\footnote{\url{http://www.iasf-milano.inaf.it/~polletta/templates/swire_templates.html}}\citep{2007ApJ...663...81P} as a typical continuum template for all galaxies (note that only the FIR part is used). For all above radiation, we take into account the redshift distortions produced by peculiar motions along the radial direction.

We also generate noise maps at each frequency by adopting\footnote{ We consider the third octile of precipitable water vapour, pwv = 0.913~mm, namely the one assumed by the ALMA Sensitivity Calculator (ASC) in the default case. The reader may also refer to Fig. 2.14 of the ALMA handbook (\url{https://almascience.eso.org/documents-and-tools/cycle-0/alma-technical-handbook/at_download/file}) } $T_{\rm sys} = 150$~K, $N_{\rm pix} = 128\times128$, and a total integration time $t_{\rm obs}=5000$~hr. The frequency range that we are considering, [238 - 316]~GHz, is sufficiently far from the prominent  water atmospheric absorption at 325 GHz. Although there is a deep decreasing trend of transmission with increasing frequency, we note that the transmission is larger than 0.9 at $\nu_0 < 300$~GHz, for pwv=1 mm, meaning that the [\CII] signal from $z > 5.3$ galaxies is not strongly affected by transmission issues. In case of other atmospheric absorption features, we could simply drop the corresponding frequency bins. This treatment would not strongly affect our conclusions.

\begin{figure*}
\centering{
\includegraphics[width=0.49\textwidth]{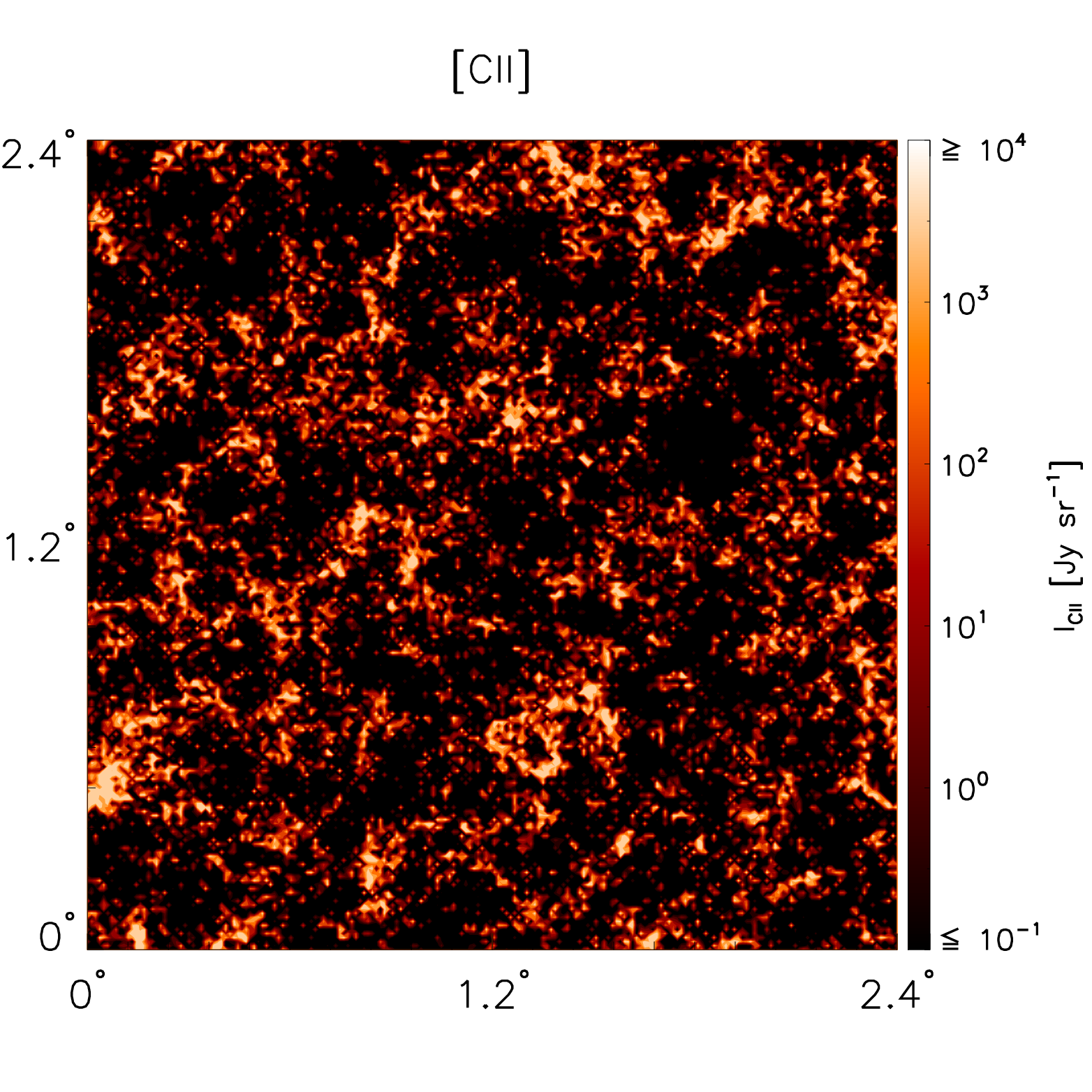}
\includegraphics[width=0.49\textwidth]{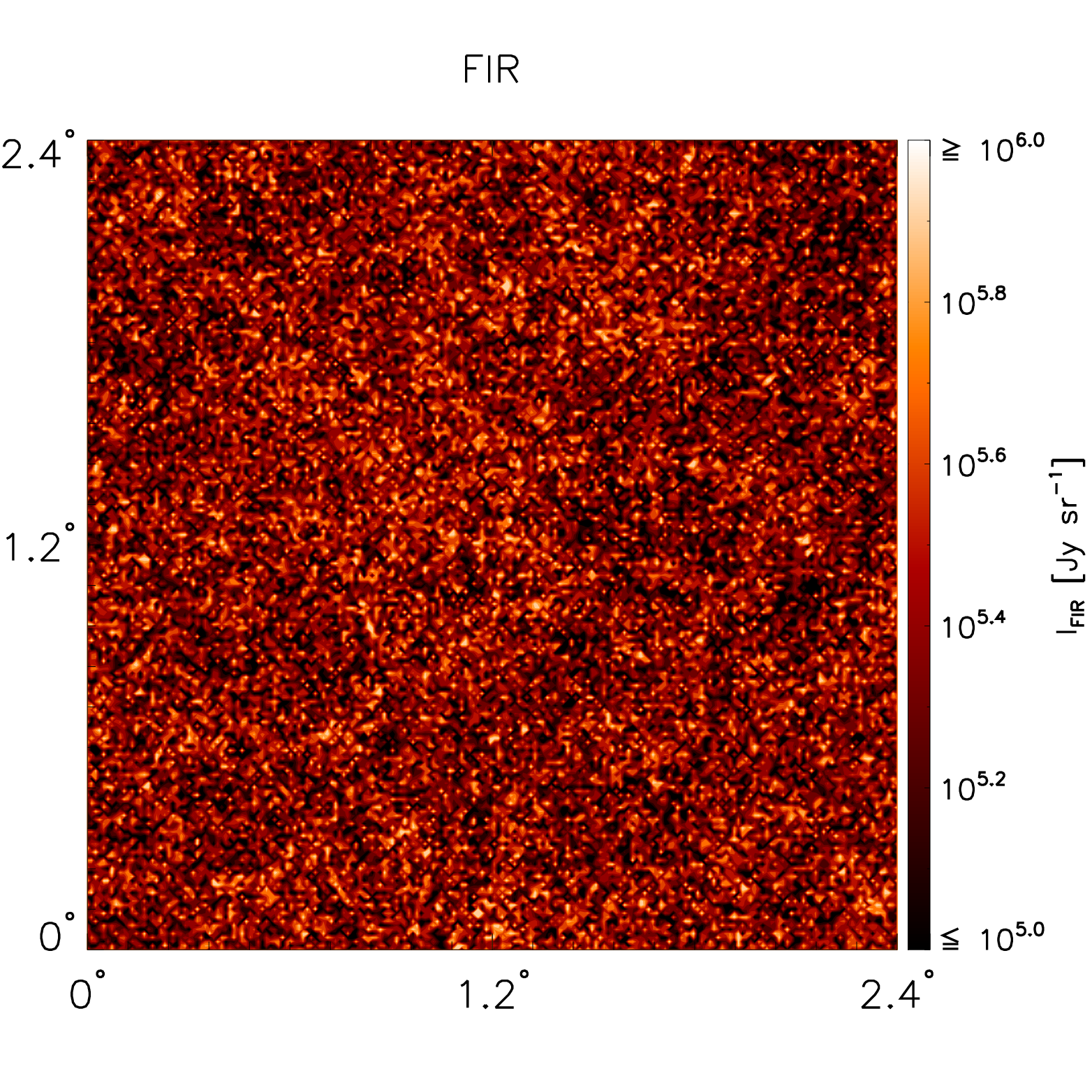} 
\includegraphics[width=0.49\textwidth]{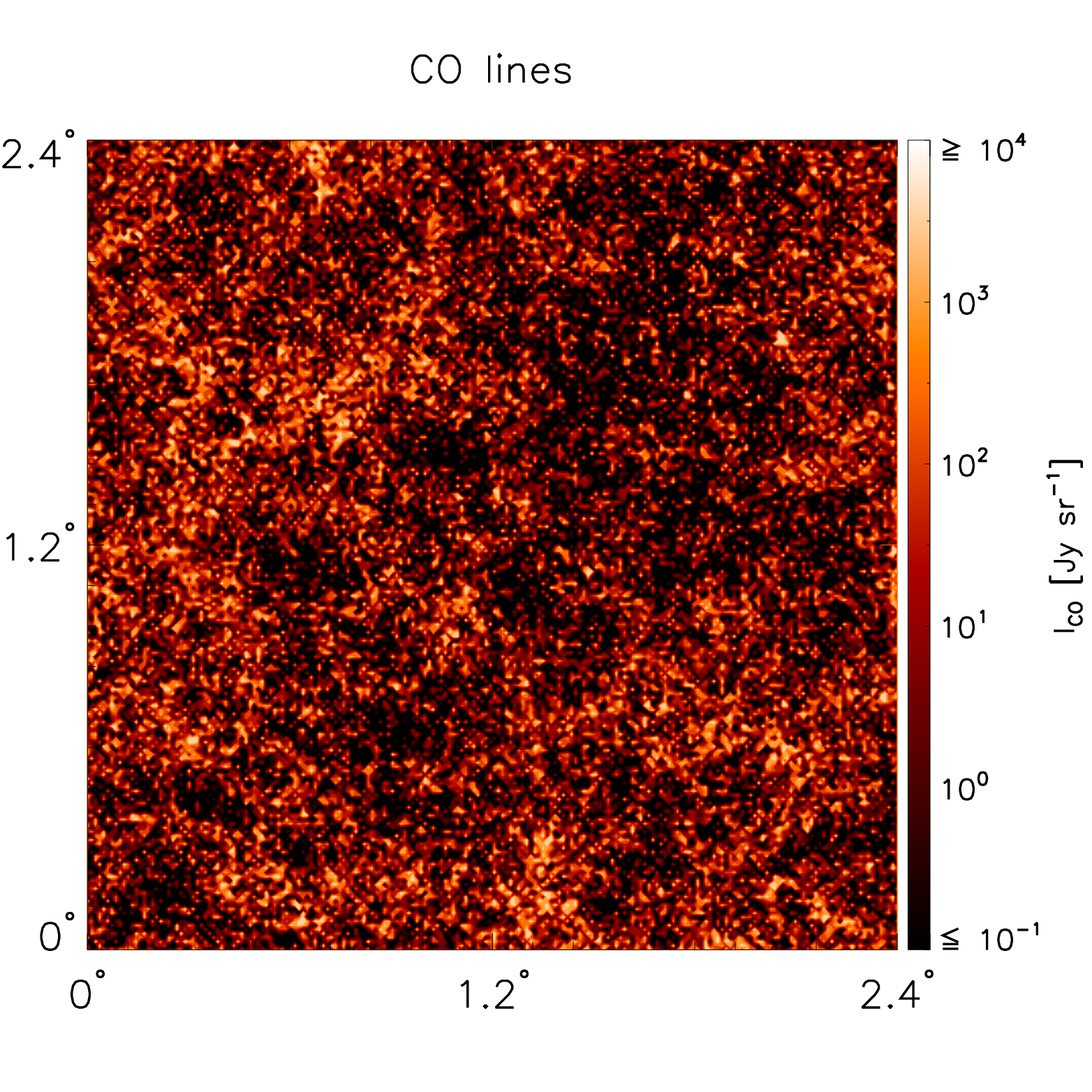}
\includegraphics[width=0.49\textwidth]{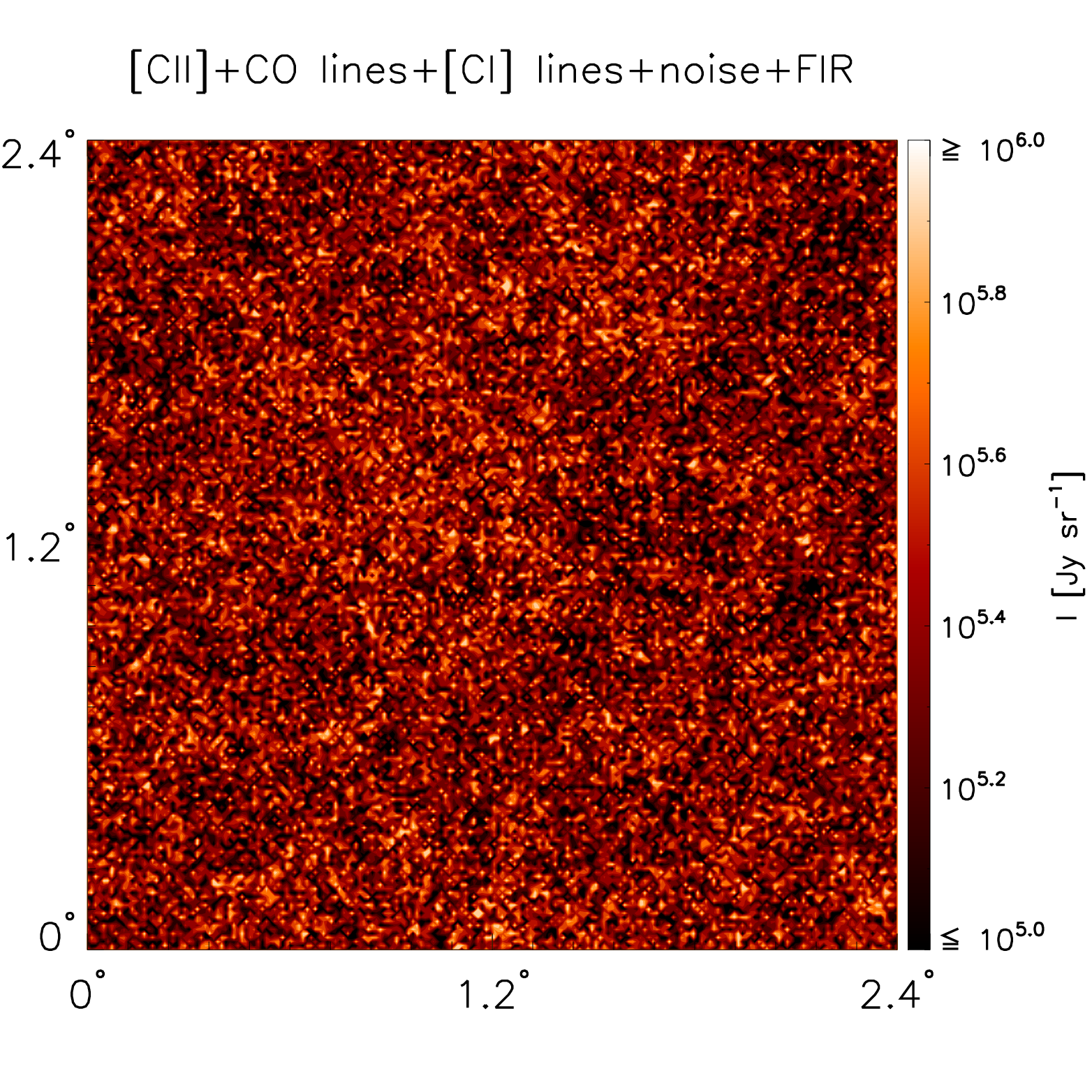}
\caption{Map of the [\CII] signal (left top), FIR continuum foreground (right top), the CO lines contamination (left bottom) and full observed map made by the sum of the signals and instrumental noise (right bottom). All maps are for the $(316\pm0.65)$~GHz frequency bin. As the emission line signal is much weaker than the continuum, the full map looks very similar to the continuum map.
}
\label{map}
}
\end{figure*}

\begin{figure}
\centering{
\includegraphics[scale=0.4]{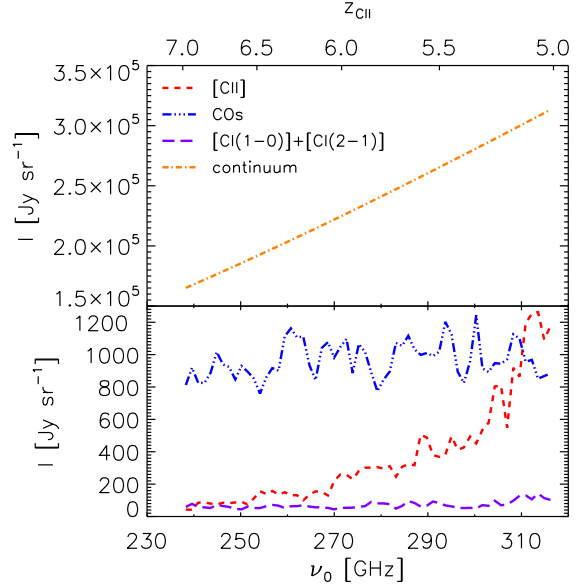}
\caption{Mean intensity of [\CII] signal, CO and [\CI] contamination and FIR continuum as a function of observed frequency. The $z_{\rm CII}$ is marked on upper abscissa. }
\label{intensity}
}
\end{figure}

For the $(316\pm 0.65)$~GHz frequency bin, maps of [\CII] signal, FIR continuum foreground, CO lines contamination and their sum plus the [\CI] lines and instrumental noise are shown separately in Fig. \ref{map}. The CO map includes all CO lines from $J = 1$ to $J = 13$. Fig. \ref{intensity} shows the mean [\CII] signal, FIR continuum, CO lines and [\CI] lines as a function of frequency. The FIR continuum is much stronger than the [\CII] and contamination emission: for instance, at 316~GHz the FIR continuum is $\sim3\times10^5$~Jy~sr$^{-1}$, while the [\CII] signal and the CO contamination are $\sim$1200~Jy~sr$^{-1}$ and $\sim$800~Jy~sr$^{-1}$, respectively. Moreover, although at $z_{\rm CII} \sim 5$, the CO lines signal is comparable to the [\CII] one, at $z_{\rm CII} \sim 7$ CO dominates by a factor $\sim20$.

At $z_{\rm CII} = 5$, the sum of the two [\CI] line fluxes is $\sim$100~Jy~sr$^{-1}$, therefore negligible with respect to the [\CII] signal. However at $z_{\rm CII} = 7$ it represents an important contamination for [\CII] intensity mapping, being  $\sim$60~Jy~sr$^{-1}$, namely comparable to the [\CII] signal.

\begin{figure}
\centering{
\includegraphics[scale=0.4]{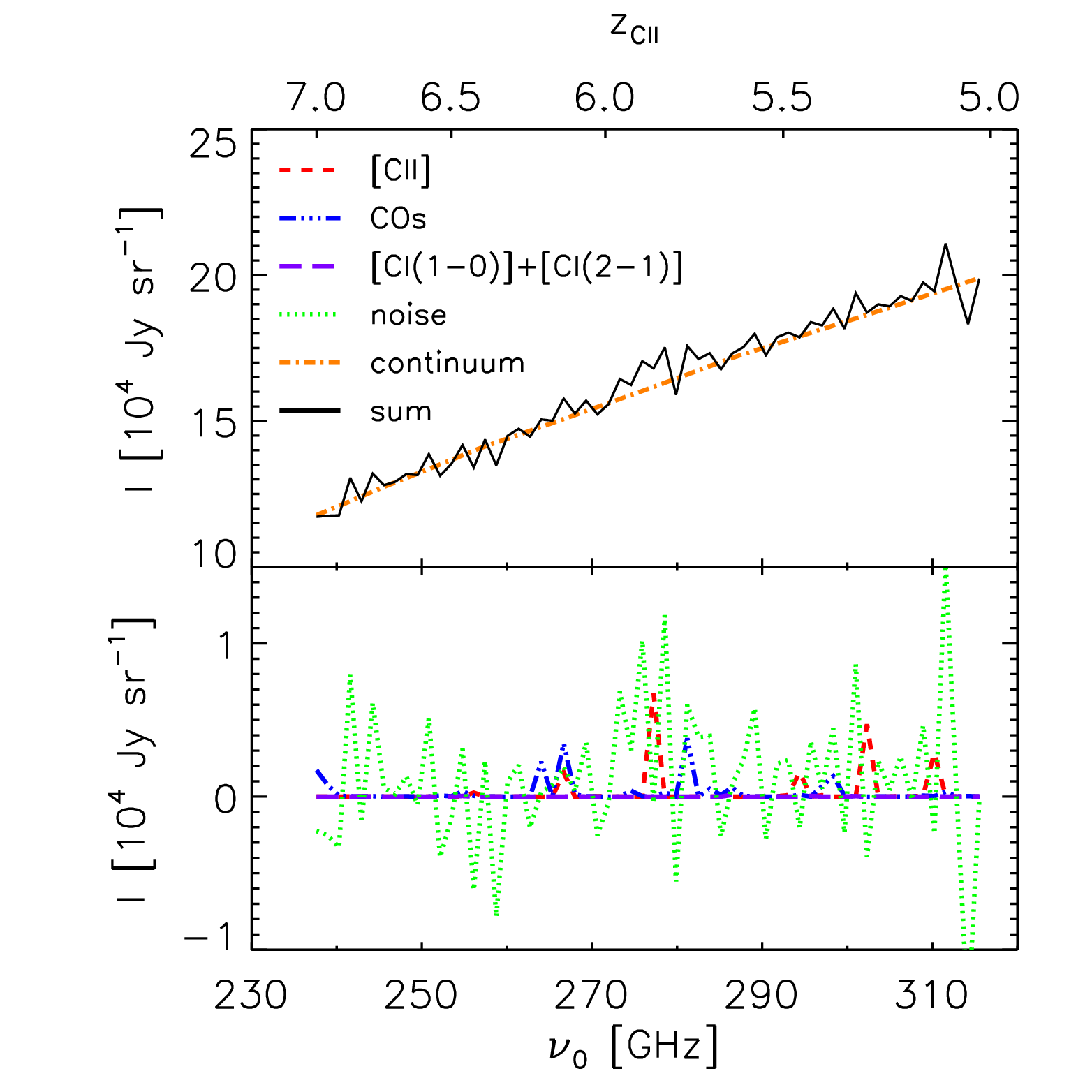}
\caption{The [\CII] signal, CO and [\CI] lines contamination, instrumental noise (bottom panel), and the FIR continuum foreground (top panel) as a function of frequency along a randomly selected line of sight cut through the mock light cone.}
\label{line}
}
\end{figure}

The frequency dependence of the fluctuations of the [\CII], FIR, CO, [\CI] and the instrumental noise, can be qualitatively appreciated by inspecting a single line of sight cut through the mock light cone, as shown in Fig. \ref{line}. 

\subsection{Recovering the [\CII] signal}\label{COmasking}

\begin{figure}
\centering{
\includegraphics[scale=0.4]{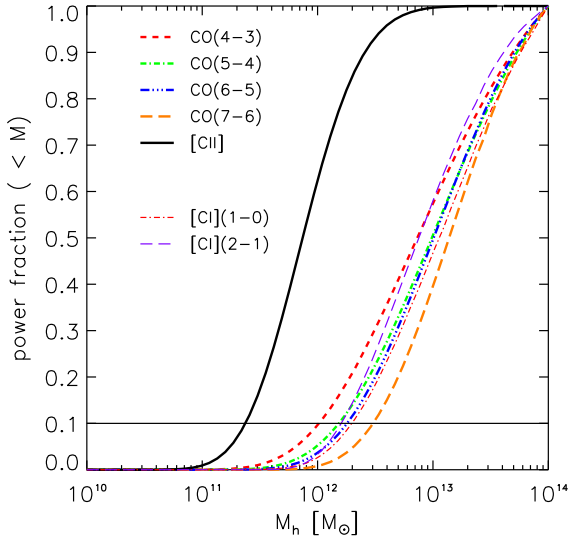}
\caption{Fractional contribution from halos below $M_{\rm h}$ to the clustering term (Eq. \ref{eq_clustering}) of the angular power spectrum for the $316\pm0.65$~GHz frequency bin for various emission lines. The two thin lines refer to [\CI(1-0)] and [\CI(2-1)] lines respectively. To guide the eye we have plot a horizontal line corresponding to 0.1.}
\label{power_fraction}
}
\end{figure}

To recover the [\CII] signal from the observed maps it is necessary to subtract the CO and [\CI] contamination and FIR continuum foreground. 
For the purpose of contamination subtraction, it is useful to know the fractional contribution from halos with different mass to the total fluctuations signal coming from galaxy clustering.
The clustering term of the angular power spectrum of a line emitted from sources located in a narrow redshift range is
\begin{equation}\label{eq_clustering}
P_{\rm clust}\propto \left[\int L(M_{\rm h})b(M_{\rm h})\frac{dn}{dM_{\rm h}}dM_{\rm h}\right]^2,
\end{equation}
where $L(M_{\rm h})$ is the line luminosity and $b(M_{\rm h})$ is the halo bias.
For $\nu_0 = 316~$GHz, the observed CO lines with $J = 4, 5, 6$ and 7 are from $z\sim 0.45, 0.82, 1.18$, and 1.54, respectively.
Most of the CO contamination to the 316~GHz map is due to these four lines. The two [\CI] lines are from $z\sim 0.56$ and 1.56 respectively. Fig. \ref{power_fraction} shows the fractional power spectrum from halos below a certain mass for [\CII] and these four CO lines and two [\CI] lines. In order to decrease CO and [\CI] contamination by $> 90\%$, radiation from halos above $\sim 10^{12} - 3\times10^{12}~M_\odot$ must be subtracted. In a $2.4\times2.4$ deg$^2$ field there are $\sim2\times10^5$ halos with $M_{\rm h}>10^{12}~M_\odot$ at all redshifts; however, only those whose CO or [\CI] emission lines are received in the frequency bin centered at $\nu_0$ produce contamination. 

In principle, de-contaminating the map would be easy if we could measure the contamination line flux of each galaxy in the map, then subtract it from the relevant pixels. However, this procedure is very time consuming. Alternatively, we could identify pixels that are supposed to be heavily contaminated by CO or [\CI] lines and discard them completely. 
We note that the CO or [\CI] sources should be much brighter than [\CII] sources, say, in the optical/IR band, where they are also more easily resolved. Thus, one can directly drop the pixels in which optical/IR bright sources are detected in the redshift range from which CO or [\CI] lines are redshifted into the observed frequency bin. Of course, this would also cause a loss of [\CII] flux from the dropped pixels. However, as [\CII] (CO or [\CI]) lines come primarily from high-$z$ (low-$z$) galaxies, there are no correlations between the two galaxy populations, i.e., removing CO([\CI])-bright pixels is equivalent to a random masking of [\CII] signal. The missing [\CII] power is limited as long as the masking fraction is $\lsim 30\%$ \citep{2005Natur.438...45K}.

How do we select optical/IR bright sources? The natural choice is to use the K-band magnitude \citep{2014arXiv1410.4808S}. However, this quantity may not be a good indicator for IR and CO or [\CI] luminosities. For example, the K-band to total IR flux ratio of the 7 spiral galaxy templates in the {\tt SWIRE} SED library varies by a factor of $\sim$30. The discrepancy for different galaxy types is even larger. To use the specific K-band brightness in contamination removal, the scatter of the $L_K - L_{\rm IR}$ relation should be modeled reasonably. 

In the samples used in \citet{2014ApJ...794..142G}, we find 51 (U)LIRGs and 15 high-$z$ DSFGs whose K-band flux can be found on their website. From such data, we fit the following $L_{K^\prime} - L_{\rm IR}$ relation:
\begin{equation}
{\rm log}\left(\frac{L_{K^\prime}}{\rm erg~s^{-1} Hz^{-1}}\right) = 0.39 \times{\rm log}\left( \frac{L_{\rm IR}}{L_\odot} \right) +25.26,
\label{LK}
\end{equation}
with a standard deviation of residuals equal to 0.35; $L_{K^\prime}$ is the luminosity at the rest frame frequency that is redshifted into K band. For a halo with mass $M_{\rm h}$ and IR luminosity $L_{\rm IR}$ (computed from Eq. \ref{LIR}),
$L_{K^\prime}$ is randomly generated from a log-normal distribution with mean given by Eq. (\ref{LK}) and standard deviation $\sigma_{K^\prime}=0.35$.
We find that, for example, a halo with typical IR luminosity $10^{11}~L_\odot$ at $z=1$ is as bright as $m_K =21$ when adopting the Spi4 spiral galaxy SED template. We are then confident that K-band luminosities can be safely used to remove CO and [\CI] contamination. The advantage with respect to relying on UV/optical magnitudes is that dust extinction effects in the K-band are much smaller, and can be neglected to a first approximation.

The FIR foreground subtraction algorithm exploits the fact that the continuum is a very smooth function in frequency space. Such feature is widely used, for example, in HI 21cm intensity mapping \citep{2006ApJ...650..529W,2008MNRAS.389.1319J,2014arXiv1409.8667A}. For this reason, we believe that assuming the same FIR continuum template for all halos is acceptable, as smoothness without specifying the slope, is the only feature of the foreground that is required by this algorithm. We check that, adopting a very different SED template, e.g. an elliptical galaxy or a starburst galaxy, result in different slopes and amplitude of predicted FIR foreground, but the recovered [\CII] signal is almost identical.

In what follows we list the steps for recovering the [\CII] signal from the full, observed map
at the 60 frequency bins in which the frequency range [238, 317]~GHz ($z_{\rm CII} = $[7, 5]) is sampled ($\Delta\nu_0 = 1.3$~GHz). They form a set of $200\times200$ line of sights (one for each pixel of the map). 
\begin{enumerate}
\item Identify the ``CO or [\CI] contaminated'' pixels (pixels containing $m_K < 22$ galaxies whose contamination lines are redshifted into the relevant band) in each line of sight, replace their flux with the interpolated value from the two neighboring pixels along the same line of sight.
\item For each line of sight, take out its foreground component that is found by either singular value decomposition (SVD), or polynomial fitting algorithm (details are given in Appendix \ref{app_foreground}). 
\item Set the flux of ``CO or [\CI] contaminated'' pixels identified in step (i) be zero. 
\end{enumerate}
After the above procedures, the final map contains the [\CII] signal, instrumental noise, and relatively negligible FIR continuum foreground and contamination residuals.
We check that, at angular scale $\sim1000''$, for frequencies corresponding to $z_{\rm CII} =5$, 6 and 7, 0.1\%, 6\% and 20\% of the CO contamination power spectrum is left as residuals, respectively. For [\CI] lines the corresponding fraction is 2\%, 5\% and 10\%, respectively.

\section{Results}\label{result}

\begin{figure}
\centering{
\includegraphics[scale=0.4]{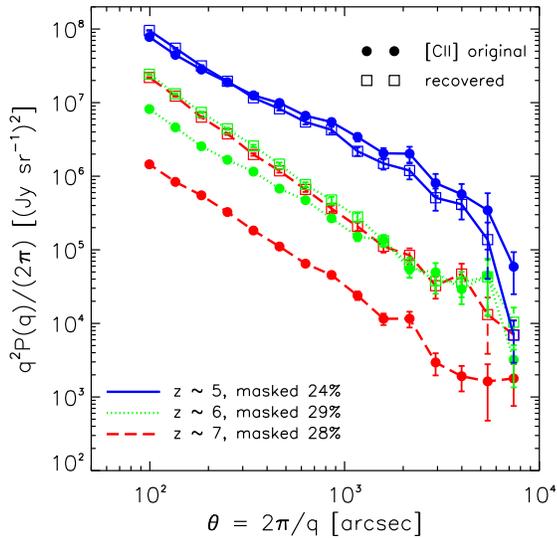}
\caption{The remaining angular power spectrum after foreground and contamination removal (open squares connected with lines) and the original [\CII] signal (filled circles connected with lines) as 
a function of angular scale $\theta = 2\pi/q$, where $q$ is the wavenumber,
at $z_{\rm CII} \sim 5$ (solid), 6 (dotted) and 7 (dashed) respectively. We give the masked percentages for each. 
The assumed observational configuration is a 6 m telescope with $T_{\rm sys} = 150$~K,
$N_{\rm pixel}=128\times128$, and a $t_{\rm obs}=5000$~hr.}
\label{angular_power_recovered}
}
\end{figure}

Fig. \ref{angular_power_recovered} shows the recovered [\CII] angular power spectrum (dashed) from $ z\sim 5, 6, 7$, along with the original signal (solid). At $z_{\rm CII}\sim5$, the power spectrum is almost perfectly recovered, with the slight deficiency at large angular scales due to the discarded [\CII] flux in CO or [\CI] contaminated pixels. At $z_{\rm CII}\sim6$, the [\CII] signal is recovered for $\theta \gsim 1000''$; at smaller scales the limiting factor is the instrumental noise that can however be suppressed by a longer integration time. The signal from $z_{\rm CII}\sim7$ remains largely inaccessible, as noise dominates at all scales.

\section{Conclusions and Discussions}\label{conclusion}

We have studied the collective [\CII] emission signal from star-forming galaxies at $z \ge 5$, as well as the influence of FIR continuum foreground and CO and [\CI] contamination on the experimental detection of the signal. 

To this aim we have combined the predicted [\CII] line luminosity as a function of the galaxy star formation rate and metallicity (derived from single galaxy simulations including a sub-grid treatment of the interstellar medium presented in V15) with a semi-empirical approach to compute the $L_{\rm CII}-M_{\rm h}$ relation. 
This relation is subsequently applied to halo catalogs built from the large-scale N-body simulation {\tt BolshoiP}, to generate mock maps of [\CII] signal. 

To compute the FIR continuum foreground, we derived the $L_{\rm IR} - M_{\rm h}$ relation via the abundance-matching technique.
As for the contamination by CO lines emitted from low-redshift galaxies, 
instead of using the poorly constrained CO luminosity functions, we use intermediate $L^{\rm J}_{\rm CO} - L_{\rm IR}$ relations that were better fitted from measurements of both local and high-redshift samples. 
We use the theoretical calculations of the [\CI] line luminosities as a function of $L_{\rm IR}$, according to the Popping et al. (2014) model, mentioned in Sec. 2.5. We generated mock maps for FIR continuum, CO and [\CI] emission, in close analogy with the [\CII] mock maps.

We carried out FIR foreground removal and contamination masking experiments on the total mock maps (containing the signal + foreground + contamination and also the instrumental noise) to recover the angular power spectrum of original [\CII] maps. We pointed out that, in order to efficiently subtract the CO and [\CI] contamination one could discard pixels that are allegedly contaminated by contamination lines. This is feasible if the map has sufficient angular resolution to avoid losing too many pixels. 
 We estimated that if the intensity map has a resolution $\sim40''$, contamination can be suppressed by dropping all pixels containing galaxies brighter than $m_K =22$ and located at the relevant redshift range.

We found that the $z > 5$ [\CII] signal comes mainly from halos in the mass range $10^{11-12} ~M_\odot$ (H-band apparent magnitude $\sim 26.8 - 23.8$); as this mass range is narrow, intensity mapping is an ideal experiment to investigate these early galaxies. The [\CII] signal from $z_{\rm CII} \sim 5 - 6$ is detectable for a ground-based, noise-limited telescope with a 6~m aperture, $T_{\rm sys} = 150$~K, a FIR camera with $128\times128$ pixels in about $5000$~hr total integration time. 
Although feasible in principle, the experiment is difficult to be performed using currently available telescopes. In addition, the integration time itself could be longer if the atmospheric conditions are on average worse than assumed here. Therefore a dedicated telescope is required and its location is essential. In any case, our study will serve as a robust guideline for the design of future facilities. 
 
A further motivation of a [\CII] intensity mapping experiment is to detect the signal from faint galaxies that are unresolved even in deepest optical/IR surveys. This is important as these faint galaxies are believed to contribute most ionizing photons to reionization \citep{2011MNRAS.414..847S,2013MNRAS.434.1486D,2007MNRAS.380L...6C,2011MNRAS.414.1455L, 2012MNRAS.420.1606J, 2012ApJ...752L...5B,2012ApJ...758...93F}. 

In what follows, we therefore discuss the feasibility of such a kind of experiment. As can be seen from Fig. \ref{power_fraction}, at $z\sim5$, faint galaxies hosted in halos below $10^{11}~M_\odot$ only produce less than 1\% of the total [\CII] power spectrum. Therefore an intensity mapping experiment aimed at detecting such faint galaxies could only be carried out by a telescope whose noise level is background-limited, such as a cryogenic space telescope. In this case the noise is mainly due to Poisson fluctuations of the CMB \citep{2010JCAP...11..016V}:
\begin{equation}
\sigma_N = \sqrt{\frac{B(T_{\rm CMB},\nu_0) h\nu}{\lambda^2_0\Delta \nu_0 t}},
\end{equation}
where $B$ is the CMB emission at $\nu_0$. For the following calculation we adopt a 2~m aperture telescope \footnote{A 2~m aperture telescope can be considered typical for a space observatory. While a larger aperture would be more helpful in reducing integration time, the main challenge is to measure the contamination flux of foreground galaxies and [\CII] flux of bright high-$z$ galaxies.} and $t_{\rm obs} = t\times \Omega_{\rm map}/\Omega_{\rm FOV}=100$~hr, which is appropriate as a reference for a space instrument with a low noise level.

To analyze the [\CII] signal of these faint galaxies, the [\CII] flux from bright galaxies needs to be measured by a high-resolution interferometer array, and then subtracted in the relevant pixels. In the $(316\pm0.65)$~GHz frequency bin, in our light-cone there are $\sim7\times10^3$ halos with [\CII] line flux $>10^{-22}$~Wm$^{-2}$. These halos have $M_{\rm h}\gsim 6\times10^{10}~M_\odot$. In addition, there are $1.3\times10^4$ halos having CO and [\CI] contamination line flux above $10^{-22}$~Wm$^{-2}$. By assuming a line width of 50~km s$^{-1}$, to resolve these halos with a signal-to-noise ratio $> 5$, the required sensitivity is $4\times10^{-5}$~Jy. For comparison, at 316~GHz for a channel width of 50~km s$^{-1}$, ALMA sensitivity with 34 antennas is $\sim 4\times10^{-5}$~Jy with 28 hr. Assuming a $\sim$(20 arcsec)$^2$ FOV for ALMA, to cover a sky region of $2.4\times2.4$ deg$^2$, we need $\sim 2\times10^5$ pointings. Therefore, even for only one frequency bin the required integration time is so high ($\sim5\times10^6$~hour!) to make the experiment unfeasible with current technology.

\begin{figure}
\centering{
\includegraphics[scale=0.4]{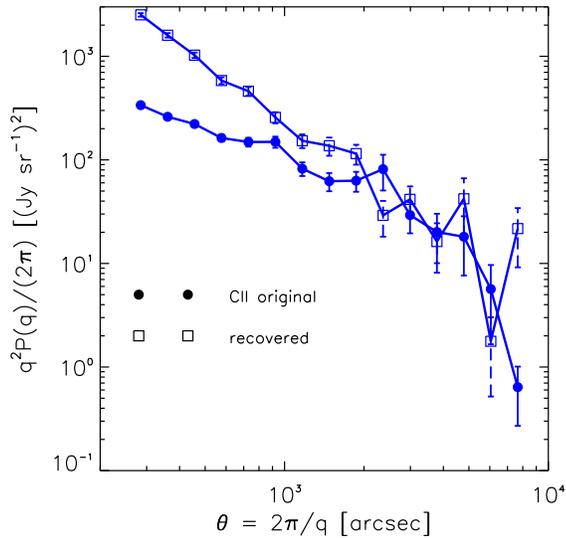}
 \caption{The original angular power spectrum of [\CII] emission from halos with [\CII] line flux $<10^{-22}$~Wm$^{-2}$ at the $316\pm 0.65$~GHz frequency bin (filled symbols) and the recovered one (open symbols).}
\label{angular_power_recovered_faint}
}
\end{figure}

In spite of this, we generate new maps including only the [\CII] signal and contamination from halos with a line flux below $< 10^{-22}$~Wm$^{-2}$. Such halos have $\lsim 10^{11}~M_\odot$, therefore we can use the $L_{\rm CII} - M_{\rm h}$ relation in P14 simulation. We can recover the [\CII] power spectrum using the procedures detailed in Sec. \ref{COmasking}, with the following considerations. Here we have assumed that galaxies with [\CII]/CO/[\CI] line flux above $10^{-22}$~Wm$^{-2}$ have already been resolved. Thus, when removing contamination, we can directly subtract the flux of the resolved sources in the relevant frequency bin, instead of discarding the pixel completely. Additionally, here the [\CII] signal to FIR continuum ratio is even smaller, which allows us to take out the first 15 modes in continuum subtraction. In Fig. \ref{angular_power_recovered_faint} we show the original and recovered power spectrum from $z_{\rm CII}\sim 5$. We can see that the experiment is indeed possible at $z=5$. At higher redshift, though, the recovery of the [\CII] signal from faint galaxies is still hampered by the noise level.

During the final stages of this work, \citet{2014arXiv1410.4808S} presented a investigation of the [\CII] signal from high-$z$ galaxies through mock surveys obtained from semi-numerical simulations. The two studies are in broad agreement, although they differ in the conclusions concerning the FIR continuum, which we found to be much stronger than both the [\CII] signal and contamination. Hence, accurately subtracting this foreground is vital in order to recover the [\CII] signal. The good news are that we showed here that the proposed algorithm to remove a spectrally smooth component from each line of sight, inspired by 21 cm experiments, can be successfully applied to [\CII] intensity mapping as well.

Finally, we comment on a possible caveat of our work, related to the fact that the V15 model neglects the contribution of HII regions to the [\CII] emission. We have already shown that HII regions in the diffuse medium do not have significant contribution compared with the cold neutral medium and warm neutral medium (see Fig. 8 in \citealt{2014ApJ...784...99G}), and V15 further shows that [\CII] emission by cold neutral medium and warm neutral medium is negligible compared with PDRs. For what concerns HII regions surrounding molecular clouds (MC), we consider the MC properties predicted by the V15 model: number density, $n_{\rm H}\sim10^{3}$~cm$^{-3}$; size, $r_{\rm MC}\sim1$~pc; ionization parameter\footnote{The quoted value for $U$ is estimated without considering the MC optical depth $\tau_{\rm MC}$ to ionizing photons. If $\tau_{\rm MC}$ is considered, $U$ will become smaller and the [\CII] emission from HII regions will further decrease.} $U \sim10^{-2.7}$. These fiducial values for $n_{\rm H}$ and $U$ are close to the parameters used in Fig. 2 of \citet{2011A&A...526A.149N}, where HII regions represent MC outskirts with column density $4\times10^{20}$~cm$^{-2}$. From this figure, we can see that in correspondence of the MC column density expected by the V15 model ($\sim 3\times10^{21}$~cm$^{-2}$) the [\CII] emission is almost 50 time stronger than the one at the edge of the HII regions, implying that the latter only contribute $f_{\rm HII}\sim2\%$ to the total emission. This calculation allows us to conclude that in the V15 model HII regions are expected to provide a negligible contribution to the total [\CII] emission, once compared with PDRs. This conclusion is consistent with the results by \citet{2006MNRAS.368.1949A}. In fact, this author finds that for $n_{\rm H}=10^3$~cm$^{-3}$ and $U=10^{-2.7}$ $f_{\rm HII}\lsim10\%$. 

The HII regions contribution to the [\CII] emission remains, however, a very controversial topic. For example, \citet{2006ApJ...652L.125O,2011ApJ...739..100O} find $f_{\rm HII}\sim(30-40)\%$ in Carina Nebula, which is a single star formation region in Milky Way. \citet{2010MNRAS.404.1910V} report fractions spanning a range $5.5\% - 60\%$ in their samples. \citet{2014ApJ...782L..17D} observed two Lyman Alpha Emitters (LAE) at $z = 4.7$, concluded that in these LAEs (which are actually members of an interacting system including quasars) most of the [\CII] emission is from the ionized medium. However, we also note that \citet{2015arXiv150203131C} found that HII regions contribution is typically less than 15\% for dwarf galaxies. 

To summarize, the contribution of HII regions to the [\CII] emission is not clearly known, especially for galaxies at $z > 5$, since [\CII] observations are still poor at these epochs. Given that $f_{\rm HII}$ is expected to be negligible according to current state-of-the-art theoretical models, we do not consider the [\CII] emission from HII regions as a major component in our calculations. Accounting for this contribution (e.g. $f_{\rm HII} \sim30\%$) would only enhance the [\CII] power spectrum (by a factor of $\sim2$), making the signal stronger.

\appendix

\section{foreground subtraction}\label{app_foreground}

Each line of sight is a 60-elements vector ${\bm A}_i$ where the smooth FIR continuum is the dominant component.
Thus, to subtract the FIR continuum it is necessary first to find its principle components, and then to subtract them from ${\bm A}_i$. In other words, expressing ${\bm A}_i$ as the linear combination of a series of base vectors that are orthogonal to each other, i.e ${\bm A}_i = \sum_j b_j {\bm B}_j$, $b_j$ indicates the contribution of each vector to the ${\bm A}_i$. With descending order of $b_j$, the summary of the first several ${\bm B}_j$ is considered as the FIR component. To find the principle components, the $60\times60$-elements covariance matrix is
$${\bm C}=\frac{1}{N}\sum_i^{N}{\bm A_i}\otimes{\bm A_i},$$
where $N=200\times200$ is the number of line of sights. By performing SVD on this matrix a series of eigenvectors sorted in descending eigenvalues $S_j$ are got, they are ${\bm B}_j$. Then $b_j$ is obtained as $b_j = {\bm A}_i\cdot {\bm B}_j$. $S_j$ descends more gently after a certain $j_{\rm max}$ ($j_{\rm max}=2$ for recovering our Fig. \ref{angular_power_recovered}, and $j_{\rm max}=15$ for recovering our Fig. \ref{angular_power_recovered_faint}), therefore most FIR continuum is considered to be represented by the first $j_{\rm max}$ principle components.

In the frequency range considered in this work, the FIR continuum is very close to a power law, thus we try the polynomial fitting algorithm as well. The FIR continuum is found to be well fitted by polynomial up to two orders and the recovered [\CII] angular spectrum is quite similar. In the main text, we have presented results obtained by adopting the SVD method. However, the polynomial fitting method would have provided consistent results.
\end{document}